\documentclass{elsarticle}
\usepackage{epsfig}
\usepackage{graphics}
\usepackage{amssymb}
\usepackage[margin=2cm]{geometry}

\usepackage{color}
\usepackage{ifpdf}
\usepackage{ifvtex}

\def\eq#1{(\ref{#1})}

\begin{document}

\date{}

\title{Entropic corrected Newton's law of gravitation and the Loop Quantum Black Hole gravitational atom}

\author[a]{R.G.L. Arag\~{a}o} 
\ead{renangomezzz@gmail.com}
\author[a]{C.A.S.Silva} 
\ead{carlos.souza@ifpb.edu.br}

\address[a]{Instituto Federal de Educa\c{c}\~{a}o Ci\^{e}ncia e Tecnologia da Para\'{i}ba (IFPB),\\ Campus Campina Grande - Rua Tranquilino Coelho Lemos, 
671, Jardim Dinam\'{e}rica
I.}

\begin{abstract}

One proposal by Verlinde \cite{Verlinde:2010hp} is that gravity is not a fundamental, 
but an entropic force. In this way, Verlinde
has provide us with a way to derive the Newton's law of gravitation from the Bekenstein-Hawking entropy-area formula.
On the other hand, since it has been demonstrated that this formula is susceptible to quantum gravity corrections,
one may hope that these corrections could be inherited by the Newton's law.
In this way, the entropic interpretation of
Newton's law could be a prolific way in order to get verifiable or
falsifiable quantum corrections to ordinary gravity in an observationally accessible regimes.
Loop quantum gravity is a theory that provide a way to approach the quantum properties of spacetime. 
From this theory, emerges a quantum corrected semiclassical black hole solution 
called loop quantum black holes or self-dual black holes. 
Among the interesting features of loop quantum black holes is the fact that they give rise to a modified 
entropy-area relation where quantum gravity corrections are present. 
In this work, 
we obtain the quantum corrected Newton's law from the entropy-area relation
given by loop quantum black holes. In order to relate our results with the recent experimental
activity, we consider the quantum mechanical properties of a huge gravitational atom 
consisting in a light neutral elementary
particle in the presence of a loop quantum black hole.\\

\end{abstract}

\maketitle

\twocolumn

\section{Introduction}

Since the rising of black hole thermodynamics, in the seventies, 
through the Hawking demonstration that all black holes
emit blackbody radiation \cite{Hawking:1974sw}, investigations about these objects break up the limits of 
astrophysics. In fact, 
black holes have been put in 
the heart of the debate of the most fascinating issues in theoretical 
physics. 
Among these issues, the search for a better understanding of the quantum nature of gravity, since 
the quantum behavior of spacetime 
must be revealed within the presence of a black hole strong gravitational field.

%%%%%%%%%%%%%%%%%%%%%%%%%%%%%%%%%%%%%%%%%%%%%%%%%%
%%%%%%%%%%%%%%%%%%%%%%%%%%%%%
Among the most important lessons from black hole thermodynamics, arises the Bekenstein-Hawking formula 
which establishes that, in a different way from other usual thermodynamical systems, the 
entropy of a black hole is not given as proportional to its volume, but to its horizon area: $S = k_{B}c^{3}A/4\hslash G$.
A deep intersection between
gravity, quantum mechanics, and thermodynamics could be contained in Bekenstein-Hawking formula, since
it
gives us one of the few situations in physics
where the Newton's gravitational constant $G$ and the speed of light $c$ meet the Planck constant $\hslash$ and the Boltzmann constant $k_{B}$.
In fact, it has been shown by String theory and
Loop Quantum Gravity that the black-hole thermodynamics must have its origin in the
atomic structure of the spacetime \cite{Rovelli:1996dv, Ashtekar:1997yu, Strominger:1996sh}.
Moreover, in \cite{Silva:2008kh, Silva:2010ir, Silva:2014jda} it has been argued that a topology change process due to the dynamics of
the quantum spacetime could be the origin of black hole entropy and the Generalized Second Law of black hole
thermodynamics.

%%%%%%%%%%%%%%%%%%%%%%%%%%%%%%%%%%%%%%%%%%%%%%%%%%%%%%%%%%%%%%%%%%%%%%%%%%%%%%%%%%%%%
%%%%%%%%%%%%%%%%%%%%%%%%%%%%%%%%%%%%%%%%%%%%%%%%%%%%%%%%%%%%%%%%%
In 1995, a surprising result by Jacobson has deepened the significance of the Bekenstein-Hawking formula. 
Assuming the proportionality between entropy and horizon area, Jacobson 
derived the Einstein's field equations
by using the fundamental Clausius relation \cite{Jacobson:1995ab}. 
The procedure behind this result is to require that the Clausius relation, $\delta Q = TdS$, associating heat, 
temperature and entropy, holds for all the local Rindler 
causal horizon through each spacetime point, with
$\delta Q$ and $T$ interpreted, respectively, as the energy flux and Unruh temperature seen by an 
accelerated observer just inside 
the horizon. In this way, the spacetime
could be viewed as a kind of gas whose entropy is given by the Bekenstein-Hawking formula, and the 
Einstein's field equation as an equation of state describing this gas. 

Following Jacobson's results, several authors have addressed the issue of the relation of Einstein's equations
and thermodynamics (For a review and a voluminous list of references see \cite{Padmanabhan:2009vy}). 
More recently, Verlinde \cite{Verlinde:2010hp} conjectured that gravity is a
non fundamental interaction but would be explained as an entropic
force. In this way, the second law
of Newton is obtained
when one tie up the
entropic force with the Unruh temperature. 
On the other hand, Newton’s law of gravitation is obtained when associating these arguments 
with the holographic principle and using the equipartition law of energy.
Verlinde's formalism, have been used in several contexts including cosmological ones 
\cite{Cai:2008ys}.

%%%%%%%%%%%%%%%%%%%%%%%%%%%%%%%%%%%%%%%%%%%%%%%%%%%%%%%%%%%%%%%%%%
%%%%%%%%%%%%%%%%%%%%%%%%%%%%%%%%%%%%%
 
On the other hand, by using the
measurement result of quantum states of ultra-cold neutron
under the Earth's gravity, Kobakhidze presented an argument in opposition to Verlinde's proposal \cite{Kobakhidze:2010mn}.
The problem pointed by Kobakhidze comes from the fact that the
entropy formula defined by Verlinde formalism, in principle,
leads to a quantum neutron mixed state. However, it disagrees with the results from the ultra-cold neutron experiment. 
Kobakhidze's criticism have been 
questioned in \cite{Chaichian:2011xc} and one resolution was suggested by Abreu et al \cite{Abreu:2013rxe}. 
This resolution can be found out by abandoning the implicit assumption in \cite{Kobakhidze:2010mn} that the
entropy on the holographic screen is additive.

%%%%%%%%%%%%%%%%%%%%%%%%%%%%%%%%%%%%%%%%%%%%%%%%%%%%%%%%%%%%%%%%%%%%%%%%
%%%%%%%%%%%%%%%%%%%%%%%%%%%%%%%%%%%%%%%%%%%%%%%

On the other hand, it is also known that, in other
contexts than Einstein's gravity, the area formula of black hole entropy may not be held.
For example, when higher order curvature term appears in some gravity theory, the area formula has to be modified
\cite{Wald:1993nt}. Modifications to Bekenstein-Hawking formula also appear when quantum gravity effects are included.
For example, when a Generalized Uncertainty Principle 
(GUP) is taken into account \cite{Medved:2004yu, Bargueno:2015tea}.
In this way, it was investigated 
modifications of the entropic force due to corrections imposed on 
the area law by quantum effects
and extra dimensions \cite{Zhang:2010hi}.
Quantum gravity corrections to Bekenstein-Hawking formula also appear in the context of 
Loop Quantum Gravity. The most popular form to these corrections appear 
as logarithmic corrections which arises due
to thermal equilibrium fluctuations and quantum fluctuations \cite{Kaul:2000kf, Meissner:2004ju, Ghosh:2004rq, Chatterjee:2003uv}. 

Another way to get quantum corrections to Bekenstein-Hawking formula, which we shall follow in this work, arises 
in the context of loop quantum black holes \cite{Modesto:2008im, Modesto:2009ve, s.hossenfelder-prd81, Alesci:2011wn, Carr:2011pr, s.hossenfelder-grqc12020412}. 
A loop black hole, also called self-dual black hole, consists
in a quantum gravity corrected Schwarzschild black hole  that appears from a simplified model of Loop Quantum Gravity.
One of the most interesting results of the loop black hole scenario is 
the resolution of the black hole singularity by the self-duality property. 
This property guarantees that an asymptotic flat region corresponding to a Planck-sized wormhole arises in the place
of the black hole singularity. The wormhole throat is described by the 
Kantowski-Sachs solution. 
The thermodynamical properties of loop black holes has been addressed
in the references \cite{Modesto:2009ve, s.hossenfelder-prd81, Alesci:2011wn, Carr:2011pr, s.hossenfelder-grqc12020412}.
Moreover, in the reference \cite{Silva:2012mt},
the thermodynamical properties of loop quantum black holes were obtained by the use of a tunneling method
with the introduction of back-reaction effects. On the other hand, in the reference 
\cite{Anacleto:2015mma}, the tunneling formalism has been applied in order to include corrections due to a 
Generalized Uncertainty Principle to loop quantum black hole's thermodynamics. Among the results related with the thermodynamics of 
loop black holes, we have a quantum corrected Bekenstein-Hawking formula for the entropy of a black hole in which 
quantum gravity ingredients have been included.

Experimental issues related with loop quantum black holes have also been addressed in the literature. In this way,
gravitational lenses effects due to this kind of black holes have been investigated in \cite{Sahu:2015dea}.
On the other hand, loop quantum black hole's quasinormal modes have been calculated in  \cite{Chen:2011zzi}, \cite{Santos:2015gja}
and \cite{Cruz:2015bcj}. In the last, axial gravitational perturbations have been taken into account.

In the present work, we shall address how the Newton's law of gravitation would be modified in the presence of 
loop quantum black holes, when quantum properties of
spacetime are taken into account. Moreover, 
motivated by the results of recent experiments, we shall consider the quantum mechanical 
system of a ‘gravitational atom’ consisting in a light neutral elementary
particle in the presence of a loop quantum black hole. In particular,
we apply the Bohr Somerfeld formalism to this system, by the use of the
modified Newton’s potential, in order to obtain its energy levels.

\section{Loop quantum black holes} \label{sdbh}

Loop quantum black holes (LQBHs) appeared at the first time 
from a simplified model of Loop Quantum Gravity(LQG) \cite{Modesto:2008im}.
The LQBH's scenario is described  by a quantum gravitationally corrected Schwarzschild metric, 
and can be written in the form

\vspace{0.5cm}

\begin{equation}
ds^{2} = - G(r)dt^{2} + F^{-1}(r)dr^{2} + H(r)d\Omega^{2} \label{self-dual-metric}
\end{equation}
\vspace{0.5cm}
\noindent with

\begin{equation}
 d\Omega^{2} = d\theta^{2} + \sin^{2}\theta d\phi^{2}\; ,
\end{equation}

\vspace{0.5cm}

\noindent where, in the equation \eq{self-dual-metric}, the metric functions are given by

\vspace{1cm}

\begin{equation}
G(r) = \frac{(r-r_{+})(r-r_{-})(r+r_{*})^2}{r^{4}+a_{0}^{2}} \; , 
\end{equation}

\begin{equation}
F(r) = \frac{(r-r_{+})(r-r_{-})r^{4}}{(r+r_{*})^{2}(r^{4}+a_{0}^{2})} \; ,
\end{equation}

\noindent and

\begin{equation}
 H(r) = r^{2} + \frac{a_{0}^{2}}{r^{2}} \; , \label{h-form}
\end{equation}

\noindent where

\begin{eqnarray*}
r_{+} = 2m \;\; ; \;\; r_{-} = 2mP^{2} \; .
\end{eqnarray*}

\noindent In this way, two horizons appears in the LQBH's scenario - 
an event horizon at $r_{+}$
and a Cauchy horizon at $r_{-}$. 

Furthermore, we have that

\begin{eqnarray}
r_{*} = &\sqrt{r_{+}r_{-}} = 2mP\;.
\end{eqnarray}
\noindent where $P$ is the polymeric function given by 
\begin{equation}
P = \frac{\sqrt{1+\epsilon^{2}} - 1}{\sqrt{1+\epsilon^{2}} +1} \; ,
\end{equation}
\noindent and

\begin{equation}
a_{0} = 
\frac{A_{min}}{8\pi}\;,
\end{equation}

%%%%%%%%%%%%%%%%%%%%%%%%%%%%%%%%%%%%%%%%%%%%%%%%%%%%%%%%%%%%%%%%%%%%%%%%%%%%5
%%%%%%%%%%%%%%%%%%
\noindent where $A_{min}$ is the minimal value of area in LQG. 

In the metric \eq{self-dual-metric}, since $g_{\theta\theta}$ is not just $r^{2}$, 
$r$ is only asymptotically the usual radial coordinate.
From the form of the function $H(r)$, one obtains a more physical radial coordinate given by

\begin{equation}
R = \sqrt{r^{2}+\frac{a_{0}^{2}}{r^{2}}} \label{phys-rad} \;.
\end{equation}

\noindent In this way, the proper circumferential distance is measured by $R$.

Moreover,
the parameter $m$ in the solution is
related to the ADM mass $M$ by 

\begin{equation}
M = m(1 + P )^{2} \;. \label{mass-rel}
\end{equation}

The equation \eq{phys-rad} reveals important aspects of the LQBH's internal structure. From this expression, we have that, as $r$ decreases from $\infty$
to $0$, $R$ first decreases from $\infty$ to $\sqrt{2 a_{0}}$ at $r= \sqrt{a_{0}}$ 
and then increases again to $\infty$. The value of $R$ associated 
with the event horizon is given by

\vspace{5mm}

\begin{equation}
R_{EH} = \sqrt{H(r_{+})} = \sqrt{(2m)^{2} + \Big(\frac{a_{0}}{2m}\Big)^{2} }\; . \label{r-horizon}
\end{equation}

\vspace{5mm}

A peculiar feature in LQBH's scenario is the property of self-duality. 
This property says that if one introduces the new coordinates 
$\tilde{r} = a_{0}/r$ and $\tilde{t} = t r_{*}^{2}/a_{0}$, with $\tilde{r}_{\pm} = a_{0}/r_{\mp}$ the metric
preserves its form. The dual radius is given by $r_{dual} = \tilde{r} = \sqrt{a_{0}}$ and corresponds to the minimal possible surface element.
Moreover, since the equation \eq{phys-rad} can be written as $R = \sqrt{r^{2}+\tilde{r}^{2}}$, 
it is clear that, in the LQBH's scenario, we have another asymptotically flat Schwazschild region in the place of the 
singularity in the limit $r\rightarrow 0$. This new region corresponds to a Planck-sized wormhole. 
Figure \eq{carter-penrose} shows the Carter-Penrose diagram for the LQBH.

\begin{figure}[htb]
 \centering % figura centralizada
 \fbox{$\includegraphics[width=8cm,height=6cm]{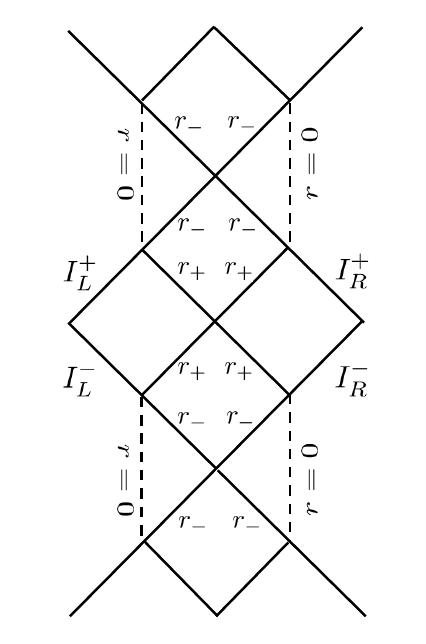}$}     
\caption[Figure 2:]{Carter - Penrose diagram for the LQBH metric. 
The diagram shows two asymptotic flat regions, one localized at infinity 
and the other near the origin, which can not be reached by an observer in a finite time. }
 \label{carter-penrose} 
\end{figure}

%%%%%%%%%%%%%%%%%%%%%%%%%%%%%%%%%%%%%%%%%%%%%%%%%%%%%%%%%%%%%%%%%%%%%%%%%%%
%%%%%%%%%%%%%%%%%%%%%%%%%%%%%%%%%%%%%%%%%%%%%%

%%%%%%%%%%%%%%%%%%%%%%%%%%%%%%%%%%%%%%%%%%%%%%%%%%%%
\vspace{2cm}

The derivation of the black hole's thermodynamical properties from the metric \eq{self-dual-metric} 
proceeds in the usual way.
The Bekenstein-Hawking temperature $T_{BH}$
can be obtained by the calculation of the surface gravity $\kappa$ by $T_{BH} = \kappa/2\pi$, with

\begin{equation}
\kappa^{2} = - g^{\mu\nu}g_{\rho\sigma}\nabla_{\mu}\chi^{\rho}\nabla_{\nu}\chi^{\sigma} = 
- \frac{1}{2}g^{\mu\nu}g_{\rho\sigma}\Gamma^{\rho}_{\mu 0} \Gamma^{\sigma}_{\nu 0}\;,
\end{equation}

\noindent where $\chi^{\mu} = (1,0,0,0)$ is a timelike Killing vector and $\Gamma^{\mu}_{\sigma\rho}$ are the connections coefficients.

By connecting with the metric, one obtains that the LQBH temperature is given by

\begin{equation}
T_{H} =  \frac{(2m)^{3}(1-P^{2})}{4\pi[(2m)^{4} +a_{0}^{2}]}\;. \label{lb-temperature}
\end{equation}

\begin{figure}[htb]
 \centering % figura centralizada
 \fbox{$\includegraphics[width=8cm,height=7cm]{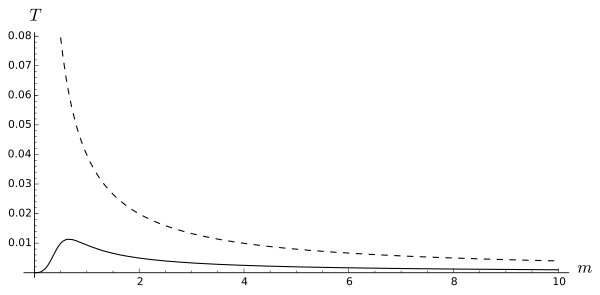}$ }    
\caption[Figure 2:]{The LQBH temperature solid line in contrast with the 
Schwarzschild black hole temperature dashed line.}
 \label{temperature-graf} 
\end{figure}

\noindent It is easy to see that one can recover the usual Hawking temperature in the limit of large masses. However, differently from
the Hawking case, the temperature \eq{lb-temperature}
goes to zero for $m \rightarrow 0$, as have been shown in the figure \eq{temperature-graf}. 
In this point, we remind that the black hole’s ADM mass
$M = m(1 + P )^2 \approx m$, since $P \ll 1$.

The black hole's entropy can be found out by making use of the
thermodynamical relation $S_{BH} = \int dm/T (m)$.

\begin{equation}
S = \frac{4\pi(1+P)^{2}}{(1-P^{2})}\Big[\frac{16m^{4} - a_{0}^{2}}{16m^{2}}\Big]\;.
\end{equation}

Moreover, the black hole entropy can be expressed in terms of its area \cite{s.hossenfelder-grqc12020412}

\begin{equation}
S =  \pm \frac{\sqrt{A^{2} - A_{min}^{2}}}{4}\frac{(1+P)}{(1-P)} \label{entropy-area}\; ,
\end{equation}

\noindent where we have set the possible additional constant to zero. S is positive for $m > \sqrt{a_{0}}/2$
and negative otherwise.

The double possibility in the signal of the loop black hole entropy is related with
the two possible physical situations that arise from LQBH structure 
\cite{Carr:2011pr}. In the first of these possibilities, 
the event horizon stays outside the wormhole throat. In order to have this situation, 
the condition $r_{+} > \sqrt{a_{0}}$ is necessary, which implies that 
$m > \sqrt{a_{0}}/2$.
In this case,
the bounce takes place after the black hole forms for a super-Planckian LQBH and the exterior,
is similar, in a qualitative way, 
to that would be produced by a Schwarzschild black hole with the same
mass. In this way, outside the event
horizon, the LQBH scenario is different from the Schwarzschild's one only by Planck-scale
corrections. On the other hand, in the sub-Planckian regime, we have a more instigating situation.
In this case, the event horizon becomes the
other side of the wormhole throat. Moreover, the deviations from the Schwarzschild metric
are very expressive and the bounce takes place
before the event horizon forms. Consequently, even 
large event horizons (which it will be for $m \ll m_{P}$)
it will be invisible to observers at $r > \sqrt{a_{0}}$.

The thermodynamics properties of LQBHs has been also obtained through the Hamilton-Jacobi
version of the tunneling formalism \cite{Silva:2012mt}. By the use of this formalism, back-reaction effects 
could be included. Moreover, extensions of the LQBH solution to scenarios where charge and 
angular momentum are preset can be found in \cite{Caravelli:2010ff}. 
The issue of information loss has been also addressed in the context of loop black holes.
In this case, it has been pointed that
the problem of information loss by black holes could be relieved in this framework \cite{Alesci:2011wn, Silva:2012mt, Alesci:2012zz}. This result may be related with the absence of
a singularity in the loop black hole interior, and consists in a forceful result in benefit of this approach.

Another interesting result in the realm of LQBH is the fact that, as have been demonstrated in \cite{Silva:2015qna},
it can been sees as the building blocks of Loop Quantum Cosmology (LQC),
in the sense that, starting from the LQBH entropy expression, LQC equations can been obtained through the use of
Jacobson formalism to obtain the Einstein's gravitational equations.

In the next sections, following the formalism developed by Verlinde \cite{Verlinde:2010hp}, 
we will derive the quantum corrected Newton's from the modified entropy-area
relation given by the equation \eq{entropy-area}.

\section{Quantum corrected Newton's law from loop quantum black holes} \label{sdbh-friedmann}

Recently, Verlinde conjectured that gravity is not fundamental but can be explained as an entropic force.
In this section, following the Verlind's entropic force approach to gravity, we shall derive quantum corrected
Newton's law of gravitation from LQBHs entropy-area relation \eq{entropy-area}.

We have that in thermodynamics, if the number of states depends on position $\Delta x$, entropic force $F$ arises
as thermodynamical conjugate of $\Delta x$. In this case, the first law of thermodynamics can be written as

\begin{equation}
F\Delta x = T\Delta S \label{force-entropy}
\end{equation}

Based on the Bekenstein's entropy bound, Verlinde postulated that when a test particle moves approaching a 
holographic screen, the change of entropy on this screen is proportional to the mass $m$ of the particle, and the distance
$\Delta x$ between the test particle and the screen

\begin{equation}
\Delta S = 2\pi k_{B}\frac{mc}{\hslash}\Delta x \label{entropy-position}
\end{equation}

\noindent to derive the entropic force hypothesis, \eq{entropy-position} should hold at least
when $\Delta x$ is smaller than or comparable with the Compton wave-length of the particle.

The temperature that appears in \eq{force-entropy} can be understood in two ways: one can relate temperature and 
acceleration using Unruh's rule

\begin{equation}
k_{B}T = \frac{1}{2\pi}\frac{\hslash a}{c} \; , \label{temp1}
\end{equation}

\noindent or relate temperature, energy and the number of used degrees of freedom using equipartition rule

\begin{equation}
E = \frac{1}{2}Nk_{B}T. \label{temp2}
\end{equation}

\noindent It is necessary to point that the temperature $T$ in equations \eq{temp1} and \eq{temp2} have different meaning.
In the first equation, the temperature is defined in the bulk. However, in the second, the temperature is defined on
the holographic screen. To admit these two temperatures to be equal is an 
further supposition in Verlinde's paper.

From the equations \eq{force-entropy}, \eq{entropy-position} and \eq{temp1} one obtains the second Newtons's 
law $F = ma$. In order to obtain the Newton's law of gravitation, one must have a way to relate the number of
bits on the holographic screen with the black hole horizon area. Assuming that the number of bits on the screen is
proportional to the horizon entropy, from the equation \eq{entropy-area}, 
we shall assume that this relation is given by 

\begin{equation}
N = \frac{(1+P)}{(1-P)}\frac{\sqrt{A^{2} - A^{2}_{min}}}{L_{P}^{2}} \label{n-relation}
\end{equation}

In this way, from the equation above together with \eq{force-entropy}, \eq{temp2} and $E = Mc^{2}$, we shall have

\begin{equation}
F = -\frac{GMm}{R^{2}}\frac{(1+P)}{(1-P)}\times\frac{1}{\sqrt{1- A^{2}_{min}/16\pi^{2} R^{4}}} \; . \label{newton-law}
\end{equation}

Moreover, for the gravitational potential $V(r) = -\int F(R)dR$, we shall have

\begin{eqnarray}
V(r) = 
-\frac{GMm}{R}\frac{(1+P)}{(1-P)}\left( 1 - \frac{A_{min}^{2}}{160\pi^{2}R^{5}} - 
\frac{A_{min}^{4}}{18432\pi^{4}R^{9}} + ...\right) \label{potential}\nonumber
\end{eqnarray}

In this way, corrections to Newton's gravitational law can be obtained from LQBH entropy-area relation.
As we can see, the deviations on the Newton's law depend on the value of minimal area $A_{min}$ in LQG, as well as
on the polymeric parameter $P$. In this way, 
the corrections found out are important in the case
of submilimeter distances, 
even though it could be realized in the context of large distances through the dependence on the 
parameter $P$.

\section{The loop quantum black hole atom}

%%%%%%%%%%%%%%%%%%%%%%%%%%%%%%%%%%%%%%%%%%%%%%%%%%%%%%%%%
%%%%%%%%%%%%%%%%%%%%%%%%%

In the seventies, Hawking introduced the possibility that 
a free charged
particle could be capture by a
primordial charged black hole forming neutral and non-relativistic ultra-heavy black
hole atoms \cite{Hawking:1971ei}. After, the term gravitational atom was coined
by V. V. Flambaum and J. C. Berengut in 2001 \cite{Flambaum:2000gf} for gravitationally bound neutral black hole and a
charged particle.

An interesting fact about gravitational atoms is that they have been pointed
as an important constituent of dark matter.
In fact, primordial black hole remnants left after the Hawking
evaporation have been considered as a source of dark matter by several authors
for more than two decades 
\cite{MacGibbon:1987my, Rajagopal:1990yx, Carr:1994ar, Adler:2001vs, Chen:2002tu, Bugaev:2008gw,
Kesden:2011ij, Lobo:2013prg, Dymnikova:2013bna, Dokuchaev:2014vda} (for a review see \cite{Carr:2005bd,
Carr:2003bj, M. Y. Khlopov}). However, a central question is whether some remnants could leave after the Hawking
evaporation, forming a stable nucleus for the gravitational atom. In other words,  
in order to have a gravitational atom system as a suitable candidate to describe dark matter, 
it would be necessary that, 
at some point of its evolution, the black hole nucleus
establish a thermal stable equilibrium with its neighborhood.

In the Schwarzschild scenario, this kind of situation is  possible for
a black hole to be in equilibrium with the CMB radiation, for a black hole mass of $4.50 \times 10^{22}kg$. 
However, this equilibrium
is not a stable one because for a Schwarzschild
black hole the temperature always increases as its mass decreases and vice versa (see the dashed line in the 
fig. \eq{temperature-graf}).

On the other hand, new phenomenon emerges in the LQBH scenario.
From equation \eq{lb-temperature}, all light enough LQBHs
would radiate until their temperature cools until the point it would be in thermal equilibrium with the CMB.
In fact a stable thermal equilibrium occurs for a black hole mass
given by $m_{\textrm{stable}} \approx 10^{-19}kg$.
Based on this feature of LQBH, Modesto et al have yet pointed to the possibility that
these objects could be an important component of dark matter \cite{Modesto:2009ve}.
In this way, one could think about the possibility of gravitational atoms where a LQBH could appear as the atomic
nucleus.

In order to give a first glance on these kind of system, let us use the expression for the gravitational force
between a LQBH and a neutral particle orbiting it given by the equation \eq{newton-law}:

\begin{eqnarray}
F = \frac{GMm}{R^{2}}\frac{(1+P)}{(1-P)}\times\frac{1}{\sqrt{1- A^{2}_{min}/16\pi^{2} R^{4}}} \nonumber \\
  = \frac{mv^{2}}{R}\; ,
\end{eqnarray}

\noindent where $v$ is the particle velocity in the orbit.

In this way, we shall have:

\begin{equation}
v = \left[\frac{(1+P)}{(1-P)}\frac{GM}{R}\right]^{\frac{1}{2}}
\times\left(1- A^{2}_{min}/16\pi^{2} R^{4}\right)^{-\frac{1}{4}} \label{velocity}
\end{equation}

Using the Bohr-Somerfeld quantization method, \\
$mvR = j\hslash$, we shall get the following equation

\begin{equation}
R^{6} - \frac{(1-P)}{(1+P)}\frac{(j\hslash)^{4}}{(GMm^{2})^{2}}R^{4} + \frac{(1-P)}{(1+P)}\frac{(j\hslash)^{4}A_{min}^{2}}{(GMm^{2})^{2}} = 0 \; ,
\end{equation}

\noindent whose only real solution is

% \begin{eqnarray*}
% R &=& \Big[\left(\frac{\hslash^{4} j^{4} A(P-1)\sqrt{27m^8A^2G^4M^4(P+1)^2-4h^8j^8(P-1)^2}}
% {2(3^{3/2})m^8G^4M^4(P+1)^2} \\
% &+&\frac{2\hslash^{12}j^{12}(P-1)^3 −27\hslash^{4}j^{4}m^{8}A^{2}G^{4}M^{4}(P-1)(P+1)^2}{54m^12G^6M^6(P+1)^3}
% \right)^{1/3} \\
% &+& \frac{\hslash^8j^8(P-1)^2}{9m^8G^4M^4(P+1)^2}\left(\frac{\hslash^4j^4A(P−1)
% \sqrt{27m^8A^2G^4M^4(P+1)^2
% −4h^8j^8(P−1)^2}{23^(3/2)m^8G^4M^4(P+1)^2}\Big]^{1/2} \label{r-solution}
% \end{eqnarray*}

\begin{eqnarray*}
&&\hspace{-1cm} R_{j} = \Big[\frac{−\hslash^4j^4(P-1)}{3m^4G^2M^2(P+1)}+\Big(\hslash^{4} j^{4} A_{min}(P-1)\times \\
&&\hspace{-1cm}\frac{\sqrt{27m^8A_{min}^2G^4M^4(P+1)^2-4h^8j^8(P-1)^2}}
{2(3^{3/2})m^8G^4M^4(P+1)^2} +\\
&&\hspace{-1cm}\frac{\hslash^{4}j^{4}(1-P)[2\hslash^{8}j^{8}(1-P)^2 
-27m^{8}A_{min}^{2}G^{4}M^{4}(P+1)^2]}{54m^{12}G^6M^6(P+1)^3}\Big)^{1/3} \\
&&\hspace{-1cm}+\frac{\hslash^{12}j^{12}(P-1)^3}{9m^8G^4M^4(P+1)^2} \times\\
&&\hspace{-1cm}\Big(\frac{
\sqrt{27m^8A_{min}^2G^4M^4(P+1)^2
-4h^8j^8(P-1)^2}}{2(3^{3/2})m^8G^4M^4(P+1)^2}\\
&&\hspace{-1cm}-\frac{2\hslash^{8}j^{8}(P-1)^2 -27m^8A_{min}^2G^4M^4(P+1)^2}{54m^{12}G^6M^6(P+1)^3}\Big)^{1/3}
\Big]^{1/2} \label{r-solution}
\end{eqnarray*}

\noindent and can be expanded as

\begin{eqnarray}
R_{j} &=& \frac{\sqrt{(P+1)(1-P)}\hslash^{2}j^{2}}{m^{2}GM(P+1)}\nonumber \\ 
&-&\frac{\sqrt{(P+1)(1-P)}m^{6}G^{3}M^{3}(P+1)}{
2\hslash^{6}j^{6}(P-1)^{2}}A_{min}^{2} \nonumber \\ 
&+& \cdots \label{r-solution}
\end{eqnarray}

\noindent where the first term corresponds to the usual gravitational atom radius, unless the $P$ parameter factors.

The energy levels $E_{j}$ of the LQBH gravitational atom are obtained from the expressions \eq{potential}, \eq{velocity} and \eq{r-solution},

% \begin{eqnarray}
% &&\hspace{-1cm}E_{j} = \frac{1}{2}mv^{2} + V =  \nonumber \\
% &-&\frac{m^3G^2M^2(P+1)^{3/2}}{2\hslash^2j^2(1-P)^{3/2}} + 
% \frac{m^{11}G^{6}M^{6}(P+1)^{7/2}}{64\hslash^{10}j^{10}\pi^{2}(P-1)^{7/2}}A_{min} \nonumber \\
% &+& \frac{m^{11}G^6M^6(P+1)^{7/2}[15m^8G^4M^4(P+1)^2 - (5120\pi^2 + 128)\hslash^8j^8\pi^2(P-1)^2]}{
% 20480\hslash^{18}j^{18}\pi^4(P-1)^{11/2}}A_{min}^2 +
% \cdots
% \end{eqnarray}

\begin{eqnarray}
&&\hspace{-1cm}E_{j} = \frac{1}{2}mv^{2} + V =  \nonumber \\
&-&\frac{m^3G^2M^2(P+1)^{3/2}}{2\hslash^2j^2(1-P)^{3/2}} + 
\frac{m^{11}G^{6}M^{6}(P+1)^{7/2}}{64\hslash^{10}j^{10}\pi^{2}(P-1)^{7/2}}A_{min} \nonumber \\
&+& \frac{m^{11}G^6M^6(P+1)^{7/2}}{
20480\hslash^{18}j^{18}\pi^4(P-1)^{11/2}}A_{min}^2  \times \nonumber \\  
&[&\hspace{-0.4cm}15m^8G^4M^4(P+1)^2  
- (5120\pi^2 - 128)\hslash^8j^8\pi^2(P-1)^2] \nonumber \\ 
&+& \cdots \nonumber\\
\end{eqnarray}

\noindent where the first therm corresponds to the usual expression to the gravitational atom energy levels
(unless the dependence on the polymeric parameter),
which can be obtained in the limit where the quantum gravity corrections goes to zero.

\section{Conclusions and Remarks}

We have derived quantum corrected Newton's gravitation law from the LQBH's entropy-area relation
using the Verlinde entropic force interpretation to gravity. Our results points to some quantum deviation 
from classical Newton's law that must have a important rule in sub-millimeter distances where Newton's gravitation
theory has not been tested yet. 

Due to its self-duality property, LQBHs can have a mass lower than Planck mass.
Particularly, for $m_{\textrm{stable}} \approx 10^{-19}kg$, a LQBH would assume a stable thermal equilibrium with
the CMB, which makes possible that this kind of black holes can be seen as a good candidate for dark matter.
In this way, impelled by the current experimental activity, we investigate the energy spectrum of a huge
gravitational atom composed by a neutral particle orbiting a LQBH.
As have been demonstrated,
this frequency depends on the quantum gravitational corrections inherited from the LQBH metric.

\section{Acknowledgements}
The authors would like to thank to Conselho Nacional de Desenvolvimento Científico e
Tecnológico - CNPQ/Brazil for the financial support.

.

\expandafter\ifx\csname url\endcsname\relax \global\long\def\url#1{\texttt{#1}}
\fi \expandafter\ifx\csname urlprefix\endcsname\relax\global\long\def\urlprefix{URL }
\fi


\begin{thebibliography}{30}

 \bibitem{Verlinde:2010hp}
  E.~P.~Verlinde,
  ``On the Origin of Gravity and the Laws of Newton,''
  JHEP {\bf 1104} (2011) 029
  [arXiv:1001.0785 [hep-th]].

 

\bibitem{Hawking:1974sw}
  S.~W.~Hawking,
  ``Particle Creation by Black Holes,''
  Commun.\ Math.\ Phys.\  {\bf 43} (1975) 199
   [Commun.\ Math.\ Phys.\  {\bf 46} (1976) 206].
  doi:10.1007/BF02345020
 


 \bibitem{Rovelli:1996dv} 
  C.~Rovelli,
  ``Black hole entropy from loop quantum gravity,''
  Phys.\ Rev.\ Lett.\  77 (1996) 3288
  [gr-qc/9603063].
  
  \bibitem{Ashtekar:1997yu}
  A.~Ashtekar, J.~Baez, A.~Corichi and K.~Krasnov,
  ``Quantum geometry and black hole entropy,''
  Phys.\ Rev.\ Lett.\  80 (1998) 904
  [gr-qc/9710007].
  
  \bibitem{Strominger:1996sh}
  A.~Strominger and C.~Vafa,
  ``Microscopic origin of the Bekenstein-Hawking entropy,''
  Phys.\ Lett.\ B 379 (1996) 99
  [hep-th/9601029].
  
  
\bibitem{Silva:2008kh}
  C.~A.~S.~Silva,
  ``Fuzzy spaces topology change as a possible solution to the black hole information loss paradox,''
  Phys.\ Lett.\ B {\bf 677} (2009) 318
  doi:10.1016/j.physletb.2009.05.031
  [arXiv:0812.3171 [gr-qc]].
\bibitem{Silva:2010ir}
  C.~A.~S.~Silva and R.~R.~Landim,
  ``A note on black hole entropy, area spectrum, and evaporation,''
  Europhys.\ Lett.\  {\bf 96} (2011) 10007
  doi:10.1209/0295-5075/96/10007
  [arXiv:1003.3679 [gr-qc]].
\bibitem{Silva:2014jda}
  C.~A.~S.~Silva and R.~R.~Landim,
  ``Fuzzy spaces topology change and BH thermodynamics,''
  J.\ Phys.\ Conf.\ Ser.\  {\bf 490} (2014) 012012.
  doi:10.1088/1742-6596/490/1/012012
  
   \bibitem{Jacobson:1995ab}
  T.~Jacobson,
  ``Thermodynamics of space-time: The Einstein equation of state,''
  Phys.\ Rev.\ Lett.\  75 (1995) 1260
  [gr-qc/9504004].
  
\bibitem{Padmanabhan:2009vy}
  T.~Padmanabhan,
  ``Thermodynamical Aspects of Gravity: New insights,''
  Rept.\ Prog.\ Phys.\  {\bf 73} (2010) 046901.


\bibitem{Cai:2008ys}
  R.~G.~Cai, L.~M.~Cao and Y.~P.~Hu,
  ``Corrected Entropy-Area Relation and Modified Friedmann Equations,''
  JHEP 0808 (2008) 090
  [arXiv:0807.1232 [hep-th]].
  
\bibitem{Kobakhidze:2010mn}
  A.~Kobakhidze,
  ``Gravity is not an entropic force,''
  Phys.\ Rev.\ D {\bf 83} (2011) 021502
  doi:10.1103/PhysRevD.83.021502
  [arXiv:1009.5414 [hep-th]].
  
\bibitem{Chaichian:2011xc}
  M.~Chaichian, M.~Oksanen and A.~Tureanu,
  ``On gravity as an entropic force,''
  Phys.\ Lett.\ B {\bf 702} (2011) 419
  doi:10.1016/j.physletb.2011.07.019
  [arXiv:1104.4650 [hep-th]].
  
\bibitem{Abreu:2013rxe}
  E.~M.~C.~Abreu and J.~A.~Neto,
  ``Considerations on Gravity as an Entropic Force and Entangled States,''
  Phys.\ Lett.\ B {\bf 727} (2013) 524
  doi:10.1016/j.physletb.2013.10.053
  [arXiv:1305.5825 [hep-th]].

\bibitem{Wald:1993nt}
  R.~M.~Wald,
  ``Black hole entropy is the Noether charge,''
  Phys.\ Rev.\ D 48 (1993) 3427
  [gr-qc/9307038].
  
 
  
   \bibitem{Medved:2004yu}
  A.~J.~M.~Medved and E.~C.~Vagenas,
  ``When conceptual worlds collide: The GUP and the BH entropy,''
  Phys.\ Rev.\ D 70 (2004) 124021
  [hep-th/0411022].
  
  \bibitem{Bargueno:2015tea}
  P.~Bargueño and E.~C.~Vagenas,
  ``Semiclassical corrections to black hole entropy and the generalized uncertainty principle,''
  Phys.\ Lett.\ B 742 (2015) 15
  
\bibitem{Zhang:2010hi}
  Y.~Zhang, Y.~Gong and Z.~H.~Zhu,
  ``Modified gravity emerging from thermodynamics and holographic principle,''
  Int.\ J.\ Mod.\ Phys.\ D {\bf 20} (2011) 1505
  doi:10.1142/S0218271811019682
  [arXiv:1001.4677 [hep-th]].
  

  


    
  \bibitem{Kaul:2000kf}
  R.~K.~Kaul and P.~Majumdar,
  ``Logarithmic correction to the Bekenstein-Hawking entropy,''
  Phys.\ Rev.\ Lett.\  84 (2000) 5255
  [gr-qc/0002040].
  
  
  \bibitem{Meissner:2004ju}
  K.~A.~Meissner,
  ``Black hole entropy in loop quantum gravity,''
  Class.\ Quant.\ Grav.\  21 (2004) 5245
  [gr-qc/0407052].
  
  \bibitem{Ghosh:2004rq}
  A.~Ghosh and P.~Mitra,
  ``A Bound on the log correction to the black hole area law,''
  Phys.\ Rev.\ D {\bf 71} (2005) 027502
  [gr-qc/0401070].
  
  \bibitem{Chatterjee:2003uv}
  A.~Chatterjee and P.~Majumdar,
  ``Universal canonical black hole entropy,''
  Phys.\ Rev.\ Lett.\  {\bf 92} (2004) 141301
  [gr-qc/0309026].
  

\bibitem{Modesto:2008im} 
  L.~Modesto,
  ``Space-Time Structure of Loop Quantum Black Hole'',
  arXiv:0811.2196 [gr-qc].
  



\bibitem{Modesto:2009ve}
  L.~Modesto and I.~Premont-Schwarz,
  ``Self-dual Black Holes in LQG: Theory and Phenomenology'',
  Phys.\ Rev.\ D 80 (2009) 064041.

\bibitem{s.hossenfelder-prd81}  

S.~Hossenfelder, L.~Modesto and I.~Premont-Schwarz,
  ``A Model for non-singular black hole collapse and evaporation'',
  Phys.\ Rev.\ D 81 (2010) 044036
  [arXiv:0912.1823 [gr-qc]].

 

\bibitem{Alesci:2011wn}
  E.~Alesci and L.~Modesto,
  ``Particle Creation by Loop Black Holes,''
  Gen.\ Rel.\ Grav.\  46 (2014) 1656
  [arXiv:1101.5792 [gr-qc]].
  
    \bibitem{Carr:2011pr}
  B.~Carr, L.~Modesto and I.~Premont-Schwarz,
  ``Generalized Uncertainty Principle and Self-dual Black Holes'',
  arXiv:1107.0708 [gr-qc].


\bibitem{s.hossenfelder-grqc12020412} 
   S.~Hossenfelder, L.~Modesto and I.~Premont-Schwarz,
  ``Emission spectra of self-dual black holes'',
  arXiv:1202.0412 [gr-qc].
  
  
  \bibitem{Silva:2012mt}
  C.~A.~S.~Silva and F.~A.~Brito,
  ``Quantum tunneling radiation from self-dual black holes,''
  Phys.\ Lett.\ B 725 (2013) 45,  456
  [arXiv:1210.4472 [physics.gen-ph]].
  
\bibitem{Anacleto:2015mma}
  M.~A.~Anacleto, F.~A.~Brito and E.~Passos,
  ``Quantum-corrected self-dual black hole entropy in tunneling formalism with GUP,''
  Phys.\ Lett.\ B {\bf 749} (2015) 181
  doi:10.1016/j.physletb.2015.07.072
  [arXiv:1504.06295 [hep-th]].
  
\bibitem{Sahu:2015dea}
  S.~Sahu, K.~Lochan and D.~Narasimha,
  ``Gravitational lensing by self-dual black holes in loop quantum gravity,''
  Phys.\ Rev.\ D {\bf 91} (2015) 063001
  doi:10.1103/PhysRevD.91.063001
  [arXiv:1502.05619 [gr-qc]].
  

  
  \bibitem{Chen:2011zzi}
  J.~H.~Chen and Y.~J.~Wang,
  ``Complex frequencies of a massless scalar field in loop quantum black hole spacetime,''
  Chin.\ Phys.\ B {\bf 20} (2011) 030401.
  doi:10.1088/1674-1056/20/3/030401
  
  \bibitem{Santos:2015gja}
  V.~Santos, R.~V.~Maluf and C.~A.~S.~Almeida,
  ``Quasinormal frequencies of self-dual black holes,''
  arXiv:1509.04306 [gr-qc].

\bibitem{Cruz:2015bcj}
M.~B.~Cruz, C.~A.~S.~Silva and F.~A.~Brito,
  ``Gravitational axial perturbations and quasinormal modes of loop quantum black holes,''
  arXiv:1511.08263 [gr-qc].


 
  
  \bibitem{Caravelli:2010ff}
  F.~Caravelli, L.~Modesto,
  ``Spinning Loop Black Holes,''
  Class.\ Quant.\ Grav.\  27 (2010) 245022.
   [arXiv:1006.0232 [gr-qc]].
  
  
  \bibitem{Alesci:2012zz}
  E.~Alesci and L.~Modesto,
  ``Hawking radiation from loop black holes'',
  J.\ Phys.\ Conf.\ Ser.\  360 (2012) 012036.
  
  
 
  
  \bibitem{Silva:2015qna}
  C.~A.~S.~Silva,
  ``On the holographic basis of Quantum Cosmology,''
  arXiv:1503.00559 [gr-qc].
    


\bibitem{Hawking:1971ei}
  S.~Hawking,
  ``Gravitationally collapsed objects of very low mass,''
  Mon.\ Not.\ Roy.\ Astron.\ Soc.\  {\bf 152} (1971) 75.

\bibitem{Flambaum:2000gf}
V.~V.~Flambaum and J.~C.~Berengut,
  ``Atom made from charged elementary black hole,''
  Phys.\ Rev.\ D {\bf 63} (2001) 084010
  doi:10.1103/PhysRevD.63.084010
  [gr-qc/0001022].
  
  
  
%%%%%%%%%%%%%%%%%%%%%%%%%%%%%%%%%%
\bibitem{MacGibbon:1987my}
  J.~H.~MacGibbon,
  ``Can Planck-mass relics of evaporating black holes close the universe?,''
  Nature {\bf 329} (1987) 308.
  doi:10.1038/329308a0
  
\bibitem{Rajagopal:1990yx} 
  K.~Rajagopal, M.~S.~Turner and F.~Wilczek,
  ``Cosmological implications of axinos,''
  Nucl.\ Phys.\ B {\bf 358} (1991) 447.
  doi:10.1016/0550-3213(91)90355-2
  
\bibitem{Carr:1994ar}
  B.~J.~Carr, J.~H.~Gilbert and J.~E.~Lidsey,
  ``Black hole relics and inflation: Limits on blue perturbation spectra,''
  Phys.\ Rev.\ D {\bf 50} (1994) 4853
  doi:10.1103/PhysRevD.50.4853
  [astro-ph/9405027].
  
\bibitem{Adler:2001vs}
  R.~J.~Adler, P.~Chen and D.~I.~Santiago,
  ``The Generalized uncertainty principle and black hole remnants,''
  Gen.\ Rel.\ Grav.\  {\bf 33} (2001) 2101
  doi:10.1023/A:1015281430411
  [gr-qc/0106080].
  
\bibitem{Chen:2002tu}
  P.~Chen and R.~J.~Adler,
  ``Black hole remnants and dark matter,''
  Nucl.\ Phys.\ Proc.\ Suppl.\  {\bf 124} (2003) 103
  doi:10.1016/S0920-5632(03)02088-7
  [gr-qc/0205106].
  
\bibitem{Bugaev:2008gw}
  E.~Bugaev and P.~Klimai,
  ``Constraints on amplitudes of curvature perturbations from primordial black holes,''
  Phys.\ Rev.\ D {\bf 79} (2009) 103511
  doi:10.1103/PhysRevD.79.103511
  [arXiv:0812.4247 [astro-ph]].
  
  
\bibitem{Kesden:2011ij}
  M.~Kesden and S.~Hanasoge,
  ``Transient solar oscillations driven by primordial black holes,''
  Phys.\ Rev.\ Lett.\  {\bf 107} (2011) 111101
  doi:10.1103/PhysRevLett.107.111101
  [arXiv:1106.0011 [astro-ph.CO]].

\bibitem{Lobo:2013prg}
  F.~S.~N.~Lobo, G.~J.~Olmo and D.~Rubiera-Garcia,
  ``Semiclassical geons as solitonic black hole remnants,''
  JCAP {\bf 1307} (2013) 011
  doi:10.1088/1475-7516/2013/07/011
  [arXiv:1306.2504 [hep-th]].
  
  
\bibitem{Dymnikova:2013bna}
  I.~Dymnikova and M.~Fil'chenkov,
  ``Graviatoms with de Sitter Interior,''
  Adv.\ High Energy Phys.\  {\bf 2013} (2013) 746894.
  doi:10.1155/2013/746894
  
\bibitem{Dokuchaev:2014vda}
  V.~I.~Dokuchaev and Y.~N.~Eroshenko,
  ``Black hole atom as a dark matter particle candidate,''
  Adv.\ High Energy Phys.\  {\bf 2014} (2014) 434539
  doi:10.1155/2014/434539
  [arXiv:1403.1375 [astro-ph.CO]].
  
\bibitem{Carr:2005bd}
  B.~J.~Carr,
  ``Primordial black holes: Recent developments,''
  eConf C {\bf 041213} (2004) 0204
  [astro-ph/0504034].
  
\bibitem{Carr:2003bj}
  B.~J.~Carr,
  ``Primordial black holes as a probe of cosmology and high energy physics,''
  Lect.\ Notes Phys.\  {\bf 631} (2003) 301
  doi:$10.1007/978-3-540-45230-0_7$
  [astro-ph/0310838].

  
\bibitem{M. Y. Khlopov}
M. Y. Khlopov and S. G. Rubin, 
Cosmological Pattern of Microphysics in Inflationary Universe, 
Kluwer Academic, Dordrecht, The Netherlands, 2004.
  


\end{thebibliography}
\end{document}